\documentstyle[12pt]{article}
\begin{document}
\def\beq{\begin{equation}}
\def\eeq{\end{equation}}
\def\bea{\begin{eqnarray}}
\def\eea{\end{eqnarray}}
\def\ve{\vert}
\def\vel{\left|}
\def\ver{\right|}
\def\nnb{\nonumber}
\def\ga{\left(}
\def\dr{\right)}
\def\aga{\left\{}
\def\adr{\right\}}
\def\rar{\rightarrow}
\def\nnb{\nonumber}
\def\la{\langle}
\def\ra{\rangle}
\def\lla{\left<}
\def\rra{\right>}
\def\ba{\begin{array}}
\def\ea{\end{array}}
\def\tep{$B \rar K \ell^+ \ell^-$}
\def\tepm{$B \rar K \mu^+ \mu^-$}
\def\tept{$B \rar K \tau^+ \tau^-$}
\def\ds{\displaystyle}


\def\lesssim{\mathrel{\mathpalette\vereq<}}
\def\vereq#1#2{\lower3pt\vbox{\baselineskip1.5pt \lineskip1.5pt
\ialign{$\m@th#1\hfill##\hfil$\crcr#2\crcr\sim\crcr}}}

\def\gtrsim{\mathrel{\mathpalette\vereq>}}

\def\alt{\lesssim}
\def\agt{\gtrsim}




\def\bos{\lower 0.5cm\hbox{{\vrule width 0pt height 1.2cm}}}
\def\boss{\lower 0.35cm\hbox{{\vrule width 0pt height 1.cm}}}
\def\aaa{\lower 0.cm\hbox{{\vrule width 0pt height .7cm}}}
\def\dol{\lower 0.4cm\hbox{{\vrule width 0pt height .5cm}}}


\title{ {\Large {\bf 
Fourth generation effects in rare exclusive 
$B \rar K^\ast \ell^+ \ell^-$ decay } } }

\author{\vspace{1cm}\\
{\small T. M. Aliev \thanks
{e-mail: taliev@metu.edu.tr}\,\,,
A. \"{O}zpineci \thanks
{e-mail: ozpineci@metu.edu.tr}\,\, and \,\,
M. Savc{\i} \thanks
{e-mail: savci@metu.edu.tr}} \\
{\small Physics Department, Middle East Technical University} \\
{\small 06531 Ankara, Turkey} }
\date{}

\begin{titlepage}
\maketitle
\thispagestyle{empty}

\begin{abstract}
Influence of the fourth generation, if ever exists, on the experimentally measurable
quantities such as invariant dilepton mass distribution, lepton forward--backward
asymmetry, and the ratio $\Gamma_L/\Gamma_T$ of the decay widths when
$K^\ast$ meson is longitudinally and transversally polarized, is studied.
Using the experimental results on the branching ratios for the 
$B \rar X_s \gamma$ and semileptonic $B \rar X_c \ell \bar \nu$ decays, the
two possible solutions of the $4\times 4$ Cabibbo--Kobayashi--Maskawa
factor $V_{t^\prime s} V_{t^\prime b}$ are obtained as a function of the
$t^\prime$--quark mass. It is observed that the results for the
above--mentioned physical quantities are essentially
different from the standard model predictions only for one solution of the
CKM factor. In this case the above--mentioned physical quantities can serve
as efficient tools in search of the fourth generation. The other solution
yields almost identical results with the SM.
\end{abstract}


\end{titlepage}

\section{Introduction}
At present Standard Model (SM) describes very successfully all low energy
experimental data. But from theoretical point of view SM is an incomplete
theory. This theory contains many unsolved open problems, such as the
origin of CP violation, mass spectrum, etc. Another one of the unsolved 
fundamental problems of SM is the number of generations. There is no any 
theoretical argument to restrict the SM to three known generations of the 
fermions. From the LEP result of the invisible partial decay width of $Z$ 
boson it follows
that the mass of the extra generation neutrino $N$ should be larger than
$45~GeV$ \cite{R1}. In this connection there comes into mind the following
question: If extra generations exist, what effect they would have in low
energy physics? This problem was studied in many works (see for example
\cite{R2}--\cite{R8} and the references therein). 

Contributions of the new generation to the electroweak radiative corrections 
were
considered in papers \cite{R10}--\cite{R16}. It was shown in \cite{R16} that
the existing electroweak data on the $Z$--boson parameters, the $W$ boson
and the top quark masses strongly excluded the existence of the new
generations with all fermions heavier than the $Z$ boson mass. However the
same data allows few extra generations, if one allows neutral leptons to
have masses close to $50~GeV$. 

The most straightforward and economical generalization of the SM to the
four--generation case is similar to the three generations present in SM
\cite{R17}, which we consider in this work. One promising area in 
experimental search of
the fourth generation, via its indirect loop effects, is the rare $B$ meson
decays. This year the upgraded $B$ factories at SLAC and KEK will provide 
us with the first experimental data. It is also well known that in the SM
$B \rar K^\ast \ell^+ \ell^-$ decay has "large" branching ratio and it
has experimentally clean signature because two leptons are present
in the final state. For this
reason this decay is one of the most probable candidates to be detected in
these machines and in our view it is the right time for an 
investigation in this direction.          

In this work we study the contribution 
of the fourth generation in the rare $B \rar K^\ast \ell^+ \ell^-$ decay. 
At the same time this decay is sensitive to the various extension of the SM, 
because it occurs only at loop level in the SM. 

New physics effects can manifest themselves through the Wilson coefficients,
whose values can be different from the ones in the SM \cite{R18,R19}, as
well as through the new operators \cite{R20}. Note that the inclusive 
$B \rar X_s \gamma$ and $B \rar X_s \ell^+ \ell^-$ decays have already been
studied with the inclusion of the fourth generation in the SM 
\cite{R6,R21,R22}.  
    
The paper is organized as follows. In section 2, we present the necessary
theoretical expressions for the $B \rar K^\ast \ell^+ \ell^-$ decay in the
SM with four generations, as well as the expressions of the other physical
observables such as forward--backward asymmetry and the ratio of the decay
widths when $K^\ast$ meson is polarized longitudinally and transversally. 
Section 3 is devoted to the numerical analysis and our conclusion.

\section{Theoretical results}

The matrix element of the $B \rar K^\ast \ell^+ \ell^-$ decay at quark level
is described by $b \rar s \ell^+ \ell^-$ transition for whom the effective
Hamiltonian at $O(\mu)$ scale can be written as
\bea
{\cal H}_{eff} &=& \frac{4 G_F}{\sqrt{2}} V_{tb}V_{ts}^\ast
\sum_{i=1}^{10} {\cal C}_i(\mu) \, {\cal O}_i(\mu)~,
\eea
where the full set of the operators ${\cal O}_i(\mu)$ and the corresponding
expressions for the Wilson coefficients ${\cal C}_i(\mu)$ in the SM 
are given in
\cite{R23,R24}. As has been noted already, in the model we consider in this
work where the fourth generation 
is introduced in the same way the three generations are introduced in
the SM, no new operators appear and clearly the full operator set is
exactly the same as in SM. The fourth generation changes only the values of
the Wilson coefficients $C_7(\mu),~C_9(\mu)$ and $C_{10}(\mu)$, via virtual
exchange of the fourth generation up quark $t^\prime$. 
The above mentioned
Wilson coefficients can be written in the following form
\bea
C_7^{tot}(\mu) &=& C_7^{SM}(\mu) + \frac{V_{t^\prime b}^\ast V_{t^\prime s}}
{V_{tb}^\ast V_{ts}} C_7^{new} (\mu) ~, \nnb \\
C_9^{tot}(\mu) &=& C_9^{SM}(\mu) + \frac{V_{t^\prime b}^\ast V_{t^\prime s}}
{V_{tb}^\ast V_{ts}} C_9^{new} (\mu) ~, \nnb \\
C_{10}^{tot}(\mu) &=& C_{10}^{SM}(\mu) + \frac{V_{t^\prime b}^\ast V_{t^\prime s}}
{V_{tb}^\ast V_{ts}} C_{10}^{new} (\mu) ~,
\eea
where the last terms in these expressions describe the contributions of the
$t^\prime$ quark to the Wilson coefficients and $V_{t^\prime b}$ and 
$V_{t^\prime s}$ are the two elements of the $4\times 4$
Cabibbo--Kobayashi--Maskawa (CKM) matrix. In deriving Eq. (2) we factored 
out the term $V_{tb}^\ast V_{ts}$ in the effective Hamiltonian given in Eq. (1). 
The explicit forms of the $C_i^{new}$ can easily be obtained from the
corresponding Wilson coefficient expressions in SM by simply substituting 
$m_t \rar m_{t^\prime}$ (see \cite{R23,R25}). Neglecting the $s$ quark mass,
the above effective Hamiltonian leads to following matrix element for the 
$b \rar s \ell^+ \ell^-$ decay
\bea
{\cal M} &=& \frac{G\alpha}{2\sqrt{2} \pi}
 V_{tb}V_{ts}^\ast
\Bigg[ C_9^{tot} \, \bar s \gamma_\mu (1-\gamma_5) b \, 
\bar \ell \gamma_\mu \ell +  
C_{10}^{tot} \bar s \gamma_\mu (1-\gamma_5) b \,
\bar \ell \gamma_\mu \gamma_5 \ell \nnb \\
&-& 2  C_7^{tot}\frac{m_b}{q^2} \bar s \sigma_{\mu\nu} q^\nu
(1+\gamma_5) b \, \bar \ell \gamma_\mu \ell \Bigg]~,     
\eea
where $q^2=(p_1+p_2)^2$ and $p_1$ and $p_2$ are the final leptons
four--momenta. The effective coefficient $C_9^{tot}$ of the operator 
${\cal O}_9=\bar s \gamma_\mu (1-\gamma_5) b \,\bar \ell \gamma_\mu \ell$
can be written in the following form
\bea
C_9^{tot} = C_9 + Y(s)~,
\eea
where $s = q^2 / m_B^2$ and the function $Y(s)$ contains the 
contributions from the one loop
matrix element of the four quark operators. A perturbative calculation leads
to the result \cite{R23,R24},
\bea
Y_{per}(s) &=& g(\hat m_c, s) (3 C_1 + C_2 + 3 C_3 + C_4 + 3 C_5
+ C_6) - \frac{1}{2} g(1, s) (4 C_3 + 4 C_4 + 3 C_5 + C_6) \nnb \\ 
&-& \frac{1}{2} g(0, s) (C_3 + 3 C_4) +
\frac{2}{9} (3 C_3 + C_4 + 3 C_5 + C_6)~,
\eea
where $\hat m_c = m_c / m_b$. 
The explicit expressions for $g(\hat m_c, s)$, $g(0, s)$, $g(1, s)$ 
and the values of $C_i$ in the SM can be found in \cite{R23,R24}.
\begin{table}[ht]
\renewcommand{\arraystretch}{1.5}
\addtolength{\arraycolsep}{3pt}
$$
\begin{array}{|c|c|c|c|c|c|c|c|c|}
\hline
C_{1} & C_{2} & C_{3} & C_{4} & C_{5} & C_{6} & C_{7}^{SM} & C_{9}^{SM} & 
C_{10}^{SM}\\ \hline
-0.248 & 1.107& 0.011& -0.026& 0.007& -0.031& -0.313& 4.344& -4.669\\ 
\hline
\end{array}
$$
\caption{The numerical values of the Wilson coefficients at $\mu = m_{b}$
scale within the SM. The corresponding
numerical value of $C^{0}$ is 0.362.}
\renewcommand{\arraystretch}{1}
\addtolength{\arraycolsep}{-3pt}
\end{table}

In addition to the short distance contribution, $Y_{per}(s)$
receives also long distance contributions,
which have their origin in the real $c\bar c$
intermediate states, i.e., $J/\psi$, $\psi^\prime$,
$\cdots$. The $J/\psi$ family is
introduced by the Breit--Wigner distribution for the resonances
through the replacement \cite{R26}--\cite{R28}
\bea
Y(s) = Y_{per}(s) + \frac{3\pi}{\alpha^2} \,
C^{(0)} \sum_{V_i=\psi_i} 
\kappa_i \, \frac{m_{V_i} \Gamma(V_i \rar \ell^+ \ell^-)}
{m_{V_i}^2 - s m_B^2 - i m_{V_i} \Gamma_{V_i}}~,
\eea     
where $C^{(0)}= 3 C_1 + C_2 + 3 C_3 + C_4 + 3 C_5 + C_6$. 
The phenomenological parameters $\kappa_i$ can be fixed from
${\cal B} (B \rar K^\ast V_i \rar K^\ast \ell^+ \ell^-) =
{\cal B} (B \rar K^\ast V_i)\, {\cal B} ( V_i \rar \ell^+ \ell^-)$,
where the data for the right hand side is given in \cite{R29}.
For the lowest resonances $J/\psi$ nad $\psi^\prime$ we will use 
$\kappa = 1.65$ and $\kappa = 2.36$, respectively. In our numerical 
analysis we use the average of  $J/\psi$ and $\psi^\prime$ 
for the higher resonances $\psi^{(i)}$ (see \cite{R30}). 

It follows from Eq. (3) that in order to calculate the decay width and
other physical observables of the exclusive $B \rar K^\ast \ell^+ \ell^-$
decay, the matrix elements 
$\lla K^\ast \vel \bar s \gamma_\mu (1 - \gamma_5) b \ver B \rra$ and 
$\lla K^\ast \vel \bar s i \sigma_{\mu\nu} q^\nu (1 + \gamma_5) b \ver \rra$ 
have to be calculated. In other words, the exclusive 
$B \rar K^\ast \ell^+ \ell^-$ decay which is described in terms of the matrix
elements of the quark operators given in Eq. (3) over meson states, can be
parametrized in terms of form factors. 
For the vector meson $K^\ast$ with polarization vector
$\varepsilon_\mu$ the semileptonic form factors of the V--A current is
defined as
\bea
\lefteqn{
\lla K^\ast(p,\varepsilon) \vel \bar s \gamma_\mu (1 - \gamma_5) b \ver
B(p_B) \rra =} \nnb \\
&&- \epsilon_{\mu\nu\rho\sigma} \varepsilon^{\ast\nu} p^\rho q^\sigma
\frac{2 V(q^2)}{m_B+m_{K^\ast}} - i \varepsilon_\mu (m_B+m_{K^\ast})
A_1(q^2) + i (p_B + p_{K^\ast})_\mu (\varepsilon^\ast q)
\frac{A_2(q^2)}{m_B+m_{K^\ast}} \nnb \\
&&+ i q_\mu \frac{2 m_{K^\ast}}{q^2} (\varepsilon^\ast q)
\left[A_3(q^2)-A_0(q^2)\right]~,
\eea
where $\varepsilon$ is the polarization vector of $K^\ast$ meson and
$q=p_B-p_{K^\ast}$ is the momentum transfer.
Using the equation of motion, the form factor $A_3(q^2)$ can be written in
terms of the form factors $A_1(q^2)$ $A_2(q^2)$ as follows
\bea
A_3 = \frac{m_B+m_{K^\ast}}{2 m_{K^\ast}} A_1 -
\frac{m_B-m_{K^\ast}}{2 m_{K^\ast}} A_2~.
\eea
In order to ensure finiteness of (8) at $q^2=0$, we demand that
$A_3(q^2=0) = A_0(q^2=0)$.
The semileptonic form factors coming from the dipole operator 
$\sigma_{\mu\nu} q^\nu (1 + \gamma_5) b$ are defined as
\bea
\lefteqn{
\lla K^\ast(p,\varepsilon) \vel \bar s i \sigma_{\mu\nu} q^\nu (1 +
\gamma_5) b \ver
B(p_B) \rra =} \nnb \\
&&4 \epsilon_{\mu\nu\rho\sigma} \varepsilon^{\ast\nu} p^\rho q^\sigma
T_1(q^2) + 2 i \left[ \varepsilon_\mu^\ast (m_B^2-m_{K^\ast}^2) -
(p_B + p_{K^\ast})_\mu (\varepsilon^\ast q) \right] T_2(q^2) \nnb \\
&&+ 2 i (\varepsilon^\ast q) \left[ q_\mu -
(p_B + p_{K^\ast})_\mu \frac{q^2}{m_B^2-m_{K^\ast}^2} \right] T_3(q^2)~.
\eea
Using the form factors, the matrix element of the 
$B \rar K^\ast \ell^+ \ell^-$ decay takes the following form
\bea
\lefteqn{
{\cal M} = \frac{G \alpha}{4 \sqrt{2} \pi} V_{tb} V_{ts}^\ast \Bigg\{
\Bigg[ (C_9^{tot}-C_{10}^{tot}) \bar \ell \gamma_\mu
(1-\gamma_5) \ell + (C_9^{tot}+C_{10}^{tot}) \bar \ell \gamma_\mu
(1+\gamma_5) \ell \bigg]} \nnb \\
&&\times\Bigg[-\epsilon_{\mu\nu\rho\sigma} \varepsilon^{\ast\nu} p_{K^\ast}^\rho
q^\sigma \frac{2 V(q^2)}{m_B+m_{K^\ast}} - i \varepsilon_\mu^\ast
(m_B+m_{K^\ast}) A_1(q^2) + i (p_B+p_{K^\ast})_\mu (\varepsilon^\ast q)
\frac{A_2(q^2)}{m_B+m_{K^\ast}} \nnb \\
&&+ i q_\mu \frac{2 m_{K^\ast}}{q^2} (\varepsilon^\ast q)
\Big[A_3(q^2) - A_0(q^2)\Big] \Bigg]
- 4 \frac{C_7^{tot}}{q^2} m_b \Bigg[ 4 \epsilon_{\mu\nu\rho\sigma}
\varepsilon^{\ast\nu} p_{K^\ast}^\rho q^\sigma T_1(q^2)\ + 
2 i \Big(\varepsilon_\mu^\ast (m_B^2-m_{K^\ast}^2)\nnb \\
&&+ (p_B+p_{K^\ast})_\mu (\varepsilon^\ast q) \Big) T_2(q^2)
+2 i (\varepsilon^\ast q) \Bigg( q_\mu - (p_B+p_{K^\ast})_\mu
\frac{q^2}{m_B^2-m_{K^\ast}^2} \Bigg) T_3(q^2) \Bigg] \bar \ell
\gamma_\mu\ell\Bigg\}~.
\eea
From Eqs. (7), (9) and (10) we observe that in calculating the physical
observables at hadronic level, i.e., for the $B \rar K^\ast \ell^+ \ell^-$
decay, we face the problem of computing the form factors. This problem is
related to the nonperturbative sector of QCD and it can be solved only in
framework a nonperturbative approach. In the present work we choose
light cone QCD sum rules method predictions for the form factors. In what follows
we will use the results of the work \cite{R31,R32,R33} in which the form 
factors are described by a three--parameter fit where the radiative 
corrections up to leading twist contribution and
SU(3)--breaking effects are taken into account. Letting
\bea
F(q^{2})\in\{V(q^2), A_{0}(q^{2}), A_{1}(q^{2}), A_{2}(q^{2}),
A_{3}(q^{2}), T_{1}(q^{2}), T_{2}(q^{2}),
T_{3}(q^{2})\}~,\nnb
\eea 
the $q^{2}$--dependence of any of these form factors could
be parametrized as \cite{R31,R32}
\bea
\label{formfac}
F(s) = \frac{F(0)}{1-a_F\,s + b_F\, s^{2}}~, \nnb
\eea
where the parameters $F(0)$, $a_F$ and $b_F$ are listed in Table 3 for each
form factor.
\begin{table}[h]
\renewcommand{\arraystretch}{1.5}
\addtolength{\arraycolsep}{3pt}
$$
\begin{array}{|l|ccc|}  
\hline
& F(0) & a_F & b_F \\ \hline
A_0^{B \rar K^*} &\phantom{-}0.47 & 1.64 & \phantom{-} 0.94 \\
A_1^{B \rar K^*} &\phantom{-}0.35 & 0.54 & -0.02 \\ 
A_2^{B \rar K^*} &\phantom{-}0.30 & 1.02 & \phantom{-} 0.08\\
V^{B \rar K^*}   &\phantom{-}0.47 & 1.50 & \phantom{-} 0.51\\
T_1^{B \rar K^*} &\phantom{-}0.19 & 1.53 & \phantom{-} 1.77\\
T_2^{B \rar K^*} &\phantom{-}0.19 & 0.36 & -0.49\\
T_3^{B \rar K^*} &\phantom{-}0.13 & 1.07 & \phantom{-} 0.16\\ \hline
\end{array}
$$
\caption{The form factors for $B\rightarrow K^\ast \ell^{+}\ell^{-}$
in a three--parameter fit \cite{R31}.}
\renewcommand{\arraystretch}{1}
\addtolength{\arraycolsep}{-3pt}
\end{table}

Using this matrix element and the helicity amplitude formalism (see for
example \cite{R34,R35,R36}), we get for the $B \rar K^\ast \ell^+ \ell^-$ 
decay width 
\newpage
\bea
\lefteqn{
\frac{d\Gamma}{dq^2 dx} = \frac{G^2 \alpha^2}{2^{14} \pi^5 m_B}
\vel V_{tb} V_{ts}^\ast \ver^2
v \lambda^{1/2}(1,r,s)} \nnb \\
&&\times \Bigg\{ 
\vel {\cal M}^{+-}_{+}\ver^2 +
\vel {\cal M}^{+-}_{-}\ver^2 +
\vel {\cal M}^{++}_{+}\ver^2 +
\vel {\cal M}^{++}_{-}\ver^2 +
\vel {\cal M}^{-+}_{+}\ver^2 +
\vel {\cal M}^{-+}_{-}\ver^2 +
\vel {\cal M}^{--}_{+}\ver^2 \nnb \\ 
&&+
\vel {\cal M}^{--}_{-}\ver^2 +
\vel {\cal M}^{++}_{0}\ver^2 +
\vel {\cal M}^{+-}_{0}\ver^2 +
\vel {\cal M}^{-+}_{0}\ver^2 +
\vel {\cal M}^{--}_{0}\ver^2 \Bigg\}~,
\eea
where superscripts denote helicities of the leptons and subscripts
correspond to the helicity of the vector meson (in our case $K^\ast$ meson).
In Eq. (11) 
\bea 
&&\lambda(1,r,s) = 1 + r^2 + s^2 - 2 r s - 2 r - 2 s~, \nnb \\
&&q^2=(p_B-p_{K^\ast})^2~, \nnb \\
&&v = \sqrt{1 - 4 m_\ell^2/q^2},~~{\rm (velocity~ of~ the~ lepton),~ and}
\nnb \\
&&x=\cos\theta,~~  (\theta = {\rm angle~ between}~ K^\ast~{\rm and}~
\ell^-), \nnb \\
&&r=m_{K^*}^2/m_B^2~. \nnb
\eea 
The explicit forms of the helicity amplitude 
${\cal M}^{\lambda_\ell \,  \lambda_\ell}_{\lambda_V}$ are as follows:
\bea
{\cal M}^{++}_{\pm} &=& \pm \sqrt{2} m_\ell \sin\theta \ga
2 C_9^{tot} H_\pm + 4 C_7^{tot} \frac{m_b}{q^2} {\cal H}_\pm \dr~,\nnb \\ \nnb \\
{\cal M}^{+-}_{\pm} &=& (-1\pm\cos\theta) \sqrt{\frac{q^2}{2}} \Bigg\{
\Big[ 2 \ga C_9^{tot} + v C_{10}^{tot}\dr \Big] H_\pm
+ 4 C_7^{tot} \frac{m_b}{q^2} {\cal H}_\pm \Bigg\}~,
\nnb \\ \nnb \\
{\cal M}^{-+}_{\pm} &=& (1\pm\cos\theta) \sqrt{\frac{q^2}{2}} \Bigg[
2 \ga C_9^{tot} - v C_{10}^{tot}\dr H_\pm
+ 4 C_7^{tot} \frac{m_b}{q^2} {\cal H}_\pm \Bigg]~, \nnb \\ \nnb \\
{\cal M}^{--}_{\pm} &=& - \ga {\cal M}^{++}_{\pm} \dr~,\nnb \\ \nnb \\
{\cal M}^{++}_{0} &=& 2 m_\ell \cos\theta \ga
2 C_9^{tot} H_0 - 4 C_7^{tot} \frac{m_b}{q^2} {\cal H}_0 \dr + 
4m_\ell C_{10}^{tot} H_S^0~, \nnb \\ \nnb \\
{\cal M}^{+-}_{0} &=& - \sqrt{q^2} \sin\theta \Bigg[
2 \ga C_9^{tot} + v C_{10}^{tot} \dr H_0 
- 4 C_7^{tot} \frac{m_b}{q^2} {\cal H}_0 \Bigg]~, \nnb \\ \nnb \\
{\cal M}^{-+}_{0} &=&  {\cal M}^{+-}_{0} (v \rar -v) ~, \nnb \\ \nnb \\
{\cal M}^{--}_{0} &=& - 2 m_\ell \cos\theta \ga         
2 C_9^{tot} H_0 - 4 C_7^{tot} \frac{m_b}{q^2} {\cal H}_0 \dr
+ 4 m_\ell C_{10}^{tot} H_S^0~,
\eea
where
\bea
H_\pm &=& m_B \Bigg[ \pm \lambda^{1/2}(1,r,s) \frac{V(q^2)}{1+\sqrt{r}} + 
(1+\sqrt{r}) A_1(q^2) \Bigg]~, \nnb \\ \nnb \\
H_0 &=& \frac{m_B}{2 \sqrt{rs}} \Bigg[
- (1-r-s) (1+\sqrt{r}) A_1(q^2) +
 \lambda(1,r,s) \frac{A_2(q^2)}{1+\sqrt{r}} \Bigg]~, \nnb \\ \nnb \\   
H_S^0 &=& - \frac{m_B \lambda^{1/2}(1,r,s)}{\sqrt{s}} A_0(q^2) \nnb \\ \nnb \\  
{\cal H}_\pm &=& 2 m_B^2\left[ \pm \lambda^{1/2}(1,r,s) T_1(q^2) +
(1-r)T_2(q^2) \right]~, \nnb \\ \nnb \\
{\cal H}_0 &=& \frac{m_B^2}{\sqrt{rs}} 
\Bigg\{ (1-r) (1-r-s) T_2(q^2)
- \lambda(1,r,s) \left[ T_2(q^2)
+ \frac{s}{1-r} T_3(q^2) \right] \Bigg\}~,
\eea

In the present paper, we study the dependence of the following measurable
physical quantities, such as \\
\hspace*{0.5cm} (i) $\Gamma_+/\Gamma_-$, \\
\hspace*{0.5cm} (ii) $\Gamma_L/\Gamma_T=\Gamma_0/(\Gamma_+ + \Gamma_-)$, \\
\hspace*{0.5cm} (iii) the lepton forward--backward asymmetry\\ 
on $q^2$ and on the $t^\prime$ quark mass for the fixed values of the "new" CKM 
factor $V_{t^\prime s}^\ast V_{t^\prime b}$.
Here the subscripts $0$, $L$ and $T$ indicate the helicities of 
the $K^\ast$ meson, respectively.
From Eq. (11), we can easily obtain the explicit
expressions for $\Gamma_+$, $\Gamma_-$ and $\Gamma_0$ as
\bea
\Gamma_\pm &=& \frac{G^2 \alpha^2}{2^{14}\pi^5 m_B}
\vel V_{tb}V_{ts}^\ast \ver^2 \int dq^2\int dx \, v \lambda^{1/2} \Bigg\{
\vel {\cal M}_\pm^{+-} \ver^2 + \vel {\cal M}_\pm^{++} \ver^2 \nnb \\
&+& \vel {\cal M}_\pm^{-+} \ver^2+\vel {\cal M}_\pm^{--} \ver^2 \Bigg\}~,
\eea
where the upper(lower) subscript in $\Gamma$ corresponds to 
${\cal M}_+({\cal M}_-)$ and
\bea
\Gamma_0 &=& \frac{G^2 \alpha^2}{2^{14}\pi^5 m_B} 
\vel V_{tb}V_{ts}^\ast \ver^2 \int dq^2\int dx \, v \lambda^{1/2} \Bigg\{
\vel {\cal M}_0^{+-} \ver^2 + \vel {\cal M}_0^{++} \ver^2 \nnb \\
&+& \vel {\cal M}_0^{-+} \ver^2+\vel {\cal M}_0^{--} \ver^2 \Bigg\}~.
\eea 

From Eqs. (14) and (15) the expressions for the ratios
$\Gamma_+/\Gamma_-$ and $\Gamma_L/\Gamma_T=\Gamma_0/(\Gamma_+ + \Gamma_-)$
can easily be obtained. These quantities are measurable from the experiments.

The normalized forward--backward asymmetry $A_{FB}$ is one of the most
useful tools in search of new physics beyond SM. Especially the determination 
of the position of the zero of $A_{FB}$ can predict about new physics
\cite{R37}.
Indeed, existence of the new physics can be confirmed by the shift in the
position of the zero of the forward--backward asymmetry. This shift of zero
position can be used in looking for new physics. Therefore in the present
work the forward--backward asymmetry $A_{FB}$ is considered, 
which defined in the following way
\bea
\frac{d}{dq^2}A_{FB}(q^2) = \frac{\displaystyle{
\int_0^1dx \frac{d\Gamma}{dq^2 dx} - \int_{-1}^0dx \frac{d\Gamma}{dq^2 dx}}}
{\displaystyle{
\int_0^1dx \frac{d\Gamma}{dq^2 dx} + \int_{-1}^0dx \frac{d\Gamma}{dq^2 dx}}}
\eea

To obtain quantitative results we need the value of the 
fourth generation CKM matrix element $\vel V_{t^\prime s}^\ast V_{tb}\ver$. For
this aim  
following \cite{R21}, we will use the experimental results of the decays
${\cal B}(B \rar X_s \gamma)$ and ${\cal B}(B \rar X_c e \bar \nu_e)$ to determine the
fourth generation CKM factor $V_{t^\prime s}^\ast V_{t^\prime b}$.
In order to reduce the uncertainties arising from $b$--quark
mass, we consider the following ratio
\bea
R = \frac{{\cal B}(B \rar X_s \gamma)}{{\cal B}(B \rar X_c e \bar \nu_e)}~. \nnb
\eea
In leading logarithmic approximation this ratio can be written as
\bea
R = \frac{\vel V_{ts}^\ast V_{tb} \ver^2}{\vel V_{cb} \ver^2} \,\,
\frac{6 \alpha \vel C_7^{tot}(m_b) \ver^2}{\pi f(\hat m_c) \kappa(\hat m_c)} ~, 
\eea
where the phase factor $f(\hat m_c)$ and ${\cal O}(\alpha_s)$ QCD correction
factor $\kappa(\hat m_c)$ \cite{R39} of $b \rar c \ell \bar \nu$ 
are given by
\bea
f(\hat m_c)&=& 1 - 8 \hat m_c^2 + 8 \hat m_c^6 - \hat m_c^8 - 
24 \hat m_c^4 \,\mbox{\rm ln} (\hat m_c^4) ~, \nnb \\ \nnb \\
\kappa(\hat m_c)&=& 1 - \frac{2 \alpha_s(m_b)}{3 \pi} 
\Bigg[ \ga \pi^2 - \frac{31}{4} \dr \ga 1 - \hat m_c \dr^2 + \frac{3}{2} 
\Bigg]~.
\eea
Solving Eq. (17) for $V_{t^\prime s}^\ast V_{t^\prime b}$ and taking into
account Eqs. (2) and (18), we get 
\bea
V_{t^\prime s}^\ast V_{t^\prime b}^\pm = \Bigg[ \pm \sqrt{
\frac{\pi R \vel V_{cb} \ver^2 f(\hat m_c) \kappa(\hat m_c)}
     {6 \alpha \vel V_{ts}^\ast V_{tb}\ver^2}} - C_7^{SM}(m_b) \Bigg]
\,\, \frac{V_{ts}^\ast V_{tb}}{C_7^{new}(m_b)}~.
\eea

It is observed from Eq. (19) that $V_{t^\prime s}^\ast V_{t^\prime b}$
depends on the $t^\prime$--quark mass and its values for different choices
of the $t^\prime$--quark mass and for the experimentally measured branching
ratio ${\cal B} (B \rar X_s \gamma ) = 3.15 \times 10^{-4}$
\cite{R40} (see also \cite{R21}), are listed in Table 3 (Here we present only
the central value. It should be noted that the LEP result on $(B \rar X_s
\gamma)$ \cite{R41} coincide with the CLEO result within the error limits
and for this reason we don't present LEP result.
\begin{table}[ht]
\renewcommand{\arraystretch}{2}
\addtolength{\arraycolsep}{5pt}
$$
\begin{array}{|c|c|c|c|c|c|c|c|}
\hline
m_{t^\prime}~(GeV) & 50 & 100 & 150 & 200 & 250 & 300 & 400 \\ \hline \hline
V_{t^\prime s}^\ast V_{t^\prime b}^{(+)}\times 10^{-2} 
&-14.48&-10.01&-8.37&-7.55&-7.07&-6.75&-6.36 \\ \hline
V_{t^\prime s}^\ast V_{t^\prime b}^{(-)}\times 10^{-3} 
&3.45&2.39&2.00&1.80&1.69&1.61&1.52 \\ \hline
\end{array}
$$
\caption{The numerical values of $V_{t^\prime s}^\ast V_{t^\prime b}$ for
different values of the $t^\prime$--quark mass. The superscripts $(+)$ and 
$(-)$ correspond to the respective signs in front of the square root in Eq.
(19)}
\renewcommand{\arraystretch}{1}
\addtolength{\arraycolsep}{-5pt}
\end{table}

From unitarity condition of the CKM matrix we have
\bea
V_{us}^\ast V_{ub} + V_{cs}^\ast V_{cb} + V_{ts}^\ast V_{tb} +
V_{t^\prime s}^\ast V_{t^\prime b} = 0 ~.
\eea
If the average values of the CKM matrix elements in the SM \cite{R29}are used,
the sum of the first three terms in Eq. (20) is about $7.6 \times 10^{-2}$.
Substituting the value of $V_{t^\prime s}^\ast V_{t^\prime b}^{(+)}$ from
Table 3, we observe that the sum of the four terms on the left--hand side of 
Eq. (19) is closer to zero compared to the SM case, since 
$V_{t^\prime s}^\ast V_{t^\prime b}$ is very close to the sum of the first
three terms, but with opposite sign. On the other if we consider 
$V_{t^\prime s}^\ast V_{t^\prime b}^{(-)}$, whose value is about $10^{-3}$
and one order of magnitude smaller compared to the previous case. However it
should be noted that the data for the CKM is not determined to a very high
accuracy, and the error in sum of first three terms in Eq. (19) is about
$\pm 0.6 \times 10^{-2}$. It is easy to see then that the value of  
$V_{t^\prime s}^\ast V_{t^\prime b}^{(-)}$ is within this error range.
In summary both $V_{t^\prime s}^\ast V_{t^\prime b}^{(+)}$ and 
$V_{t^\prime s}^\ast V_{t^\prime b}^{(-)}$ satisfy the unitarity condition
(19) of CKM. Moreover, since 
$\vel V_{t^\prime s}^\ast V_{t^\prime b}^{(-)}\ver \le 10^{-1} \times 
\vel V_{t^\prime s}^\ast V_{t^\prime b}^{(+)}\ver$,
$V_{t^\prime s}^\ast V_{t^\prime b}^{(-)}$ contribution to the physical
quantities should be practically indistinguishable from SM results,
and our numerical analysis confirms this expectation. 

\section{Numerical analysis}

Having the explicit expressions for the physically measurable quantities,
in this Section we will study the influence of the fourth generation to
these quantities.
The values of the main input parameters, which appear in the expression for
the decay widths $\Gamma_0,~\Gamma_+,~\Gamma_-$ and $A_{FB}$ are: 
\bea
m_b &=& 4.8~GeV,~~~m_c=1.35~GeV,~~~m_\tau=1.78~GeV, \nnb \\
m_\mu &=& 0.105~GeV,~~~m_B=5.28~GeV,~~~m_{K^\ast}=0.892~GeV. \nnb
\eea 

The invariant dilepton mass distribution for the $B \rar K^\ast \mu^+ \mu^-$ and 
$B \rar K^\ast \tau^+ \tau^-$ decays, with and without the long distance
effects are presented in Fig. 1 and Fig. 2, respectively, for the choice of 
$V_{t^\prime s}^\ast V_{t^\prime b}^{(-)}$, and at $m_{t^\prime} = 50~GeV,
100~GeV, 150~GeV$ and $200~GeV$. From these figures we see that the
predictions of the fourth generation model and SM on the differential 
branching ratio practically coincide in this case. This result can be
explained by the fact that $V_{t^\prime s}^\ast V_{t^\prime b}^{(-)}$ is of
the order of $10^{-3}$ and it is one order of smaller than 
$V_{ts}^\ast V_{tb}=0.038$ \cite{R29} in the SM. For this reason, the effect
of the fourth generation in this case is small. Similar conclusion for the
same case can be drawn for the lepton forward--backward asymmetry (see
Figs. (3) and (4)). However, when we consider the 
$V_{t^\prime s}^\ast V_{t^\prime b}^{(+)}$ case, the situation changes
essentially and Figs. (5) and (6) depict the invariant mass distribution for the 
$B \rar K^\ast \mu^+ \mu^-$ and $B \rar K^\ast \tau^+ \tau^-$ decays for
this case, respectively. We observe from Fig. (5) that the four generation
model predicts lower value compared to SM when $m_{t^\prime}$ is less than
$m_t$. 
At $m_{t^\prime} = 200~GeV$, the differential decay rate is enhanced almost
twice. The behavior of the differential decay width between $J/\psi$ and
$\psi^\prime$ regions differ essentially in the two considered models.
Further, we observe from Fig. (6) that, for the $B \rar K^\ast \tau^+ \tau^-$
case at $m_{t^\prime} = 150~GeV$, four generation model predicts twice as much
lower value compared to SM. As $m_{t^\prime}$ increases, departure from SM 
prediction becomes exaggerated even further. 

The dependence of the lepton forward--backward asymmetry $A_{FB}$
on $q^2$ of the 
$B \rar K^\ast \mu^+ \mu^-$ and $B \rar K^\ast \tau^+ \tau^-$ decays  are 
depicted in Figs. (7) and (8), for the 
$V_{t^\prime s}^\ast V_{t^\prime b}^{(+)}$ case. We see from Fig. (7) at 
$m_{t^\prime} = 50~GeV$, $A_{FB}$ is positive in the subinterval up to its zero
value from origin, while it is negative in the same range at all other
choices of $m_{t^\prime}$. Therefore determination of the sign of the lepton
forward--backward asymmetry in experiments can yield useful information for
establishing new physics. A careful analysis of the same figure suggests
that the shift in the position of the $A_{FB}$ for different values of 
$m_{t^\prime}$ could be an indication of the unambiguous information about
the existence of new physics \cite{R37}.

In Figs. (9) and (10) we investigate the the dependence of the ratios
$\Gamma_+/\Gamma_-$ and $\Gamma_L/\Gamma_T$ on $q^2$
for both roots of the $V_{t^\prime s}^\ast V_{t^\prime b}$,
at different values of the $m_{t^\prime}$ for the $B \rar K^\ast \mu^+
\mu^-$ and $B \rar K^\ast \tau^+ \tau^-$ decays, respectively. As have
already been noted, due to the smallness of the 
$V_{t^\prime s}^\ast V_{t^\prime b}^{(-)}$,
these ratios practically display the same behavior as predicted 
by SM for this choice, confirming our expectation. So, this case can yield no
information about new physics. On the other hand, it follows from Fig. (9)
that for the $V_{t^\prime s}^\ast V_{t^\prime b}^{(+)}$ case, the
essential departure from the SM result is quite obvious in the 
$B \rar K^\ast \mu^+ \mu^-$ decay for the ratio $\Gamma_+/\Gamma_-$. 
Moreover, the effect of the fourth generation in the ratio
$\Gamma_L/\Gamma_T$ is depicted in Fig. (10), whose behavior  
clearly shows a strong dependence on the value of the $m_{t^\prime}$, 
especially in the $B \rar K^\ast \tau^+ \tau^-$ decay.
Such a dependence can be explained as an indication of the fact that for 
$\Gamma_L/\Gamma_T$, the terms proportional to the lepton mass can give
considerable contribution. 

Finally we would like to note that $4\times 4$ CKM matrix contains three
CP--violating phases and hence CP violation might be sizeable. We will
discuss this problem elsewhere in one of the future works.

To summarize, the
exclusive rare $B \rar K^\ast \ell^+ \ell^-~(\ell=\mu,~\tau)$ 
decay has a clean experimental signature and will be
measured at the present asymmetric $B$ factories
and future hadronic HERA-B, B-TeV and LHC-B machines and 
is very sensitive to the various extensions of the Standard Model.
In the present work this decay is studied in the SM with the four generation 
model. The two solutions of the fourth generation CKM factor 
$V_{t^\prime s}^\ast V_{t^\prime b}$ have been used. 
It is found out that for the choice of the positive root of the factor
$V_{t^\prime s}^\ast V_{t^\prime b}^{(+)}$, the measurements of the
invariant mass distribution, lepton forward--backward asymmetry and the
ratio $\Gamma_+/\Gamma_-$ could be quite efficient in establishing the
fourth generation in the $B \rar K^\ast \mu^+ \mu^-$ decay, while the
measurement of the ratio $\Gamma_L/\Gamma_T$ in the 
$B \rar K^\ast \tau^+ \tau^-$ decay seems to be very informative in
searching new physics.

\newpage
\section*{Figure captions}
{\bf Fig. 1} The dependence of the invariant dilepton mass distribution for the 
$B \rar K^\ast \mu^+ \mu^-$ decay on $q^2$ at different values
of $m_{t^\prime}$, with and without the long distance effects, for the choice of
$V_{t^\prime s}^\ast V_{t^\prime b}^{(-)}$. In Figs. (1) -- (8) the ordering
of the lines with respect to different values of $m_\prime$ are the same,
as indicated in Fig. 1.\\ \\
{\bf Fig. 2} Same as Fig. 1 but for the $B \rar K^\ast \tau^+ \tau^-$
decay.\\ \\
{\bf Fig. 3} The dependence of the forward--backward asymmetry for the 
$B \rar K^\ast \mu^+ \mu^-$ decay on $q^2$ at different values
of $m_{t^\prime}$, with and without the long distance effects, for the choice of
$V_{t^\prime s}^\ast V_{t^\prime b}^{(-)}$.\\ \\
{\bf Fig. 4} Same as Fig. 3 but for the $B \rar K^\ast \tau^+ \tau^-$
decay.\\ \\
{\bf Fig. 5} Same as Fig. 1 but for the choice of 
$V_{t^\prime s}^\ast V_{t^\prime b}^{(+)}$. \\ \\
{\bf Fig. 6} Same as Fig. 2 but for the choice of 
$V_{t^\prime s}^\ast V_{t^\prime b}^{(+)}$.\\ \\
{\bf Fig. 7} Same as Fig. 3 but for the choice of 
$V_{t^\prime s}^\ast V_{t^\prime b}^{(+)}$.\\ \\
{\bf Fig. 8} Same as Fig. 4 but for the choice of 
$V_{t^\prime s}^\ast V_{t^\prime b}^{(+)}$.\\ \\
{\bf Fig. 9} The dependence of the ratio $\Gamma_+/\Gamma_-$ on 
$m_{t^\prime}$ for both roots of the 
$V_{t^\prime s}^\ast V_{t^\prime b}$, for the 
$B \rar K^\ast \mu^+ \mu^-$ and $B \rar K^\ast \tau^+ \tau^-$ decays.\\ \\
{\bf Fig. 10} The dependence of the ratio $\Gamma_L/\Gamma_T$ on 
$m_{t^\prime}$ for both roots of the 
$V_{t^\prime s}^\ast V_{t^\prime b}$, for the 
$B \rar K^\ast \mu^+ \mu^-$ and $B \rar K^\ast \tau^+ \tau^-$ decays.  

\newpage

\begin{figure}
\vskip 1.5 cm
    \includegraphics{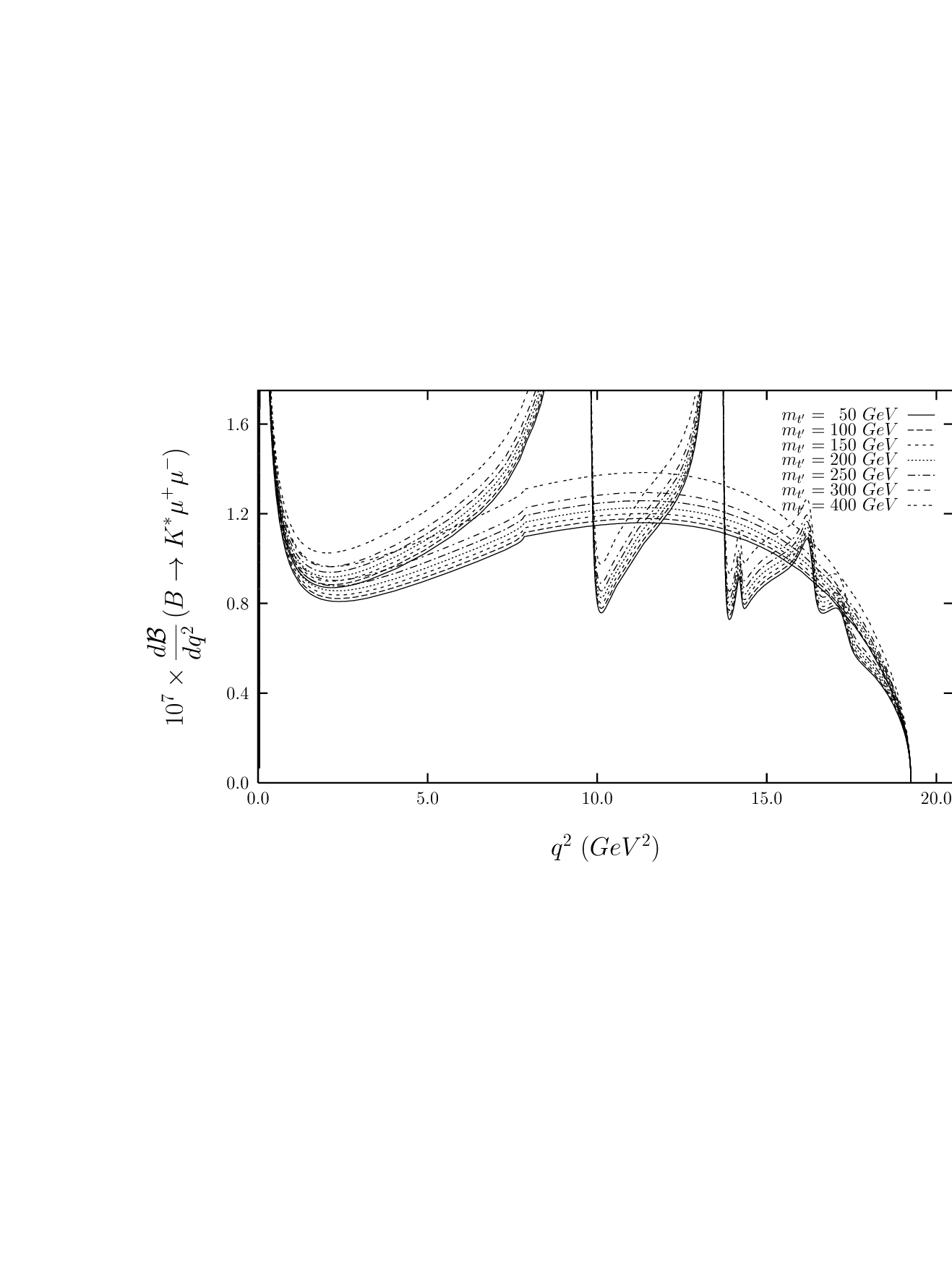}
\vskip 7.1cm
\caption{}
\end{figure}

\begin{figure}
\vskip 1.5 cm
    \includegraphics{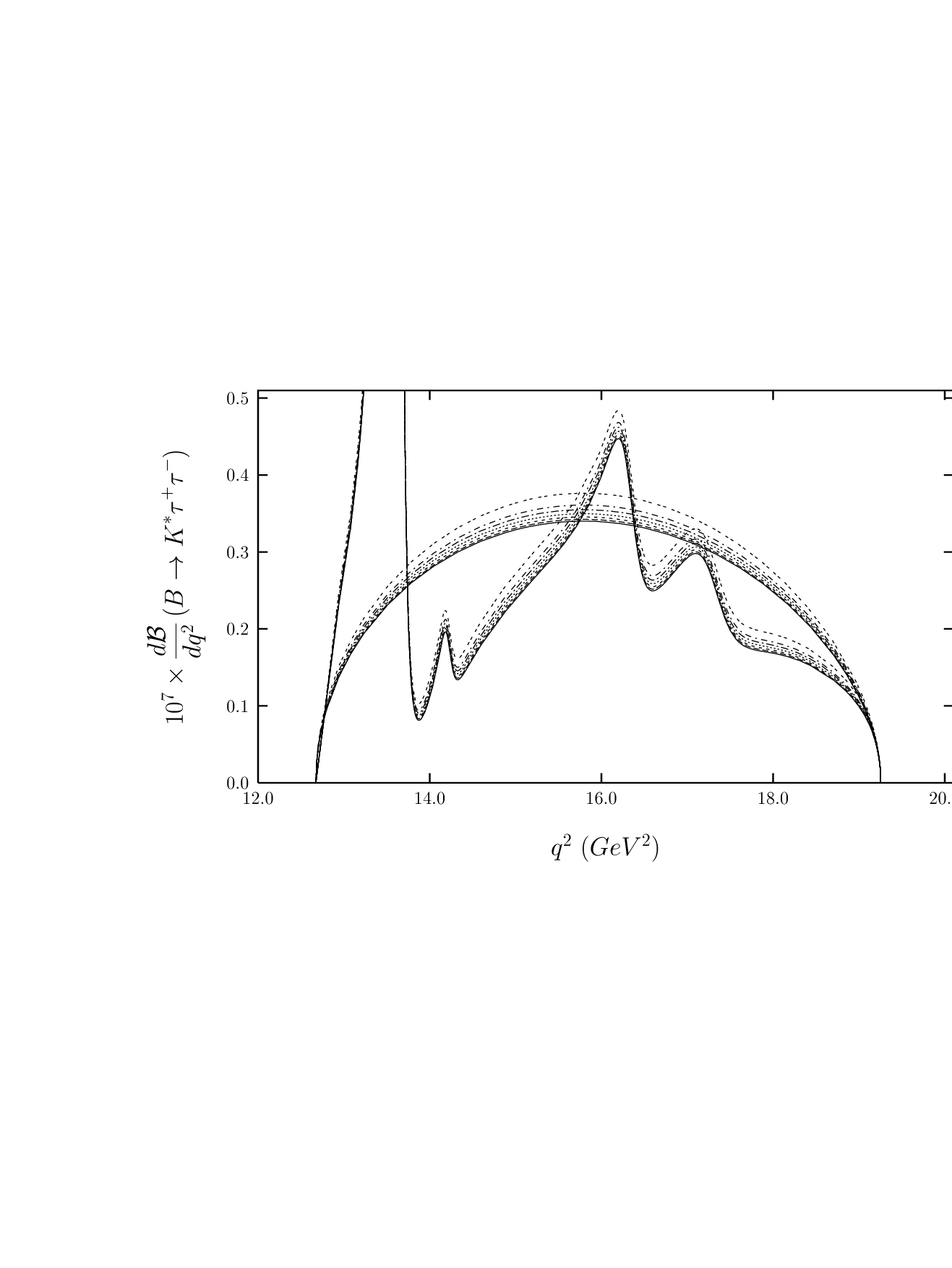}
\vskip 7.9 cm
\caption{}
\end{figure}

\begin{figure}
\vskip 1.5 cm
    \includegraphics{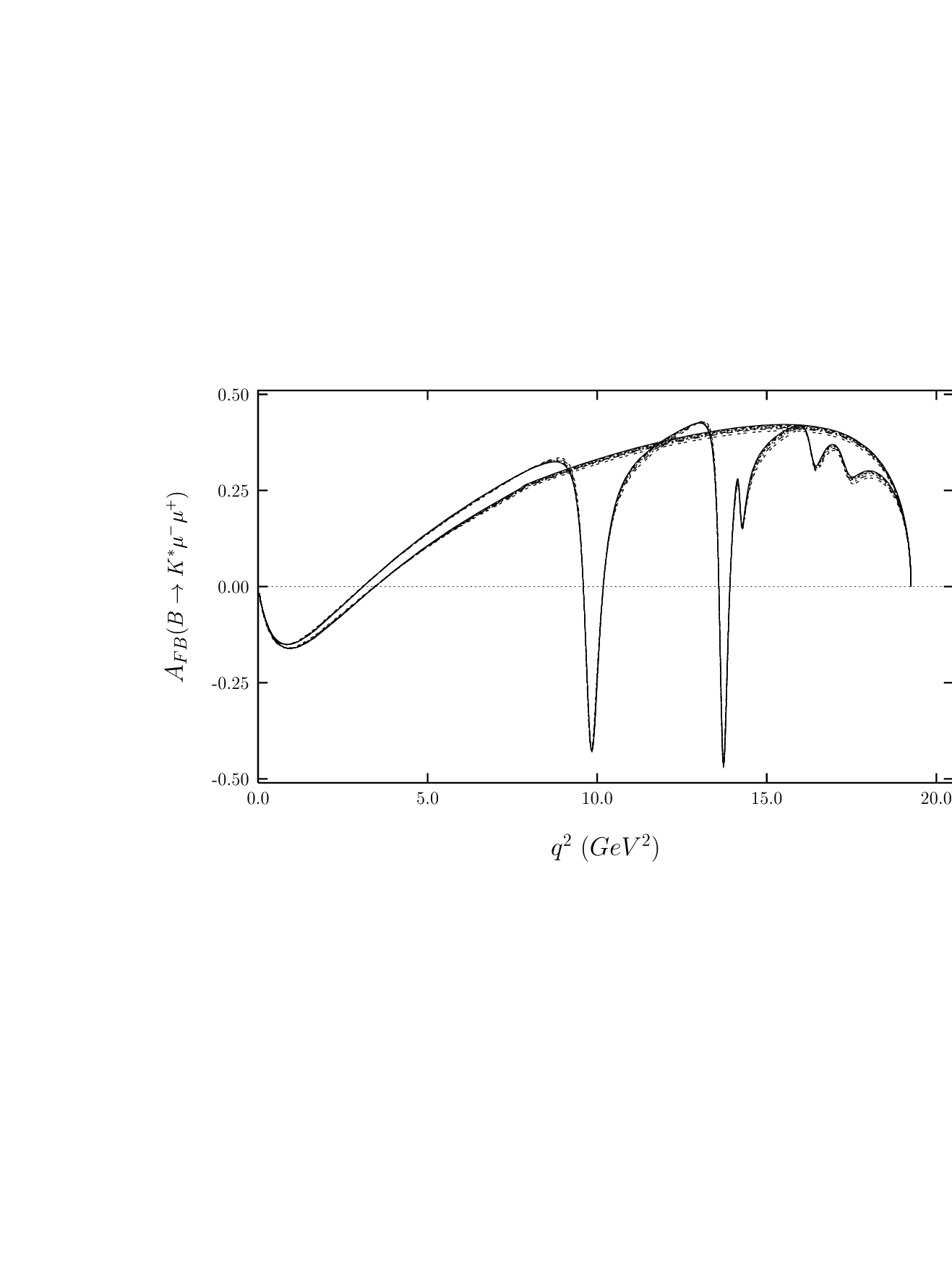}
\vskip 7.1cm
\caption{}
\end{figure}

\begin{figure}
\vskip 1.5 cm
    \includegraphics{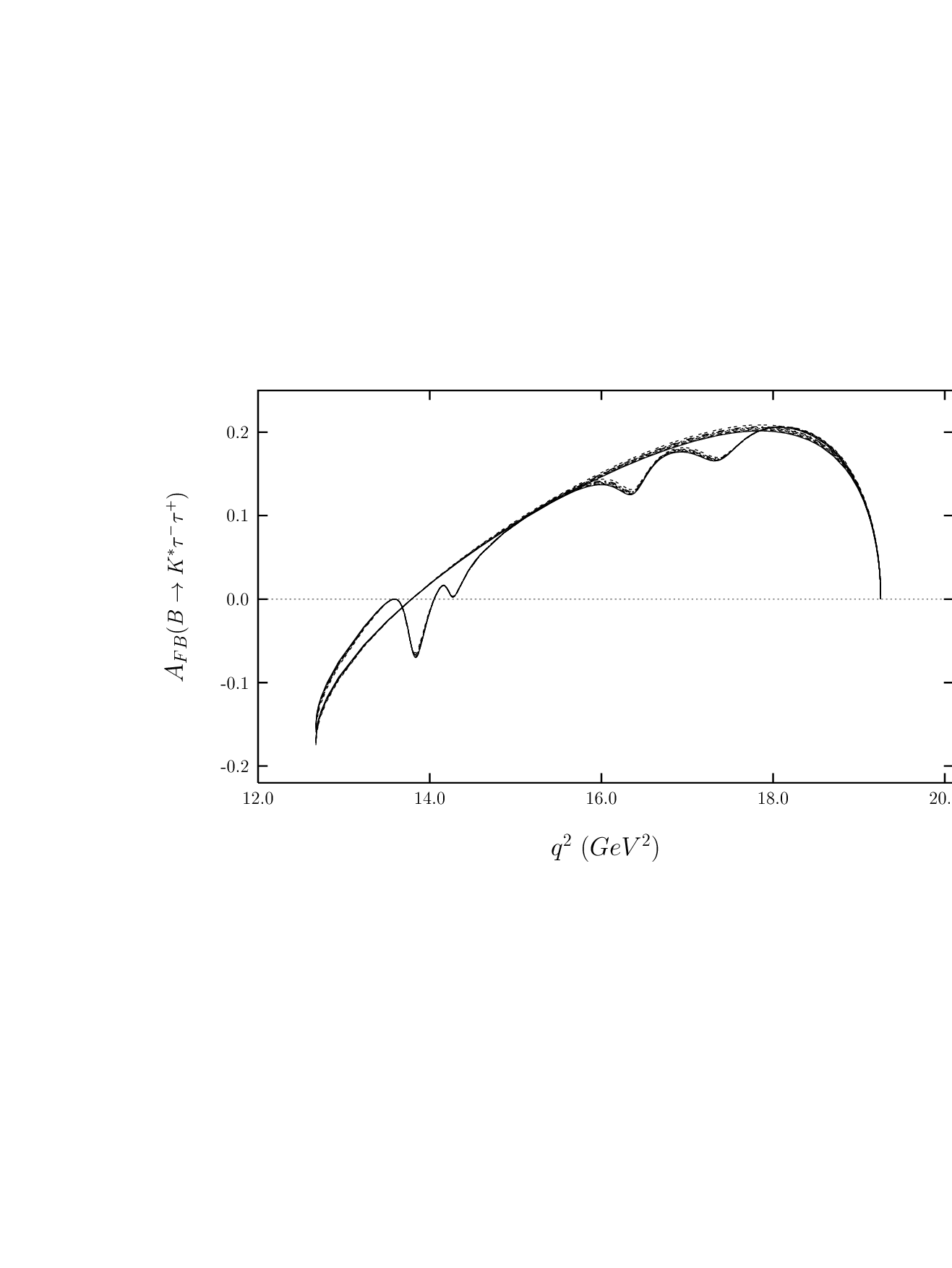}
\vskip 7.9 cm
\caption{}
\end{figure}

\begin{figure}
\vskip 1.5 cm
    \includegraphics{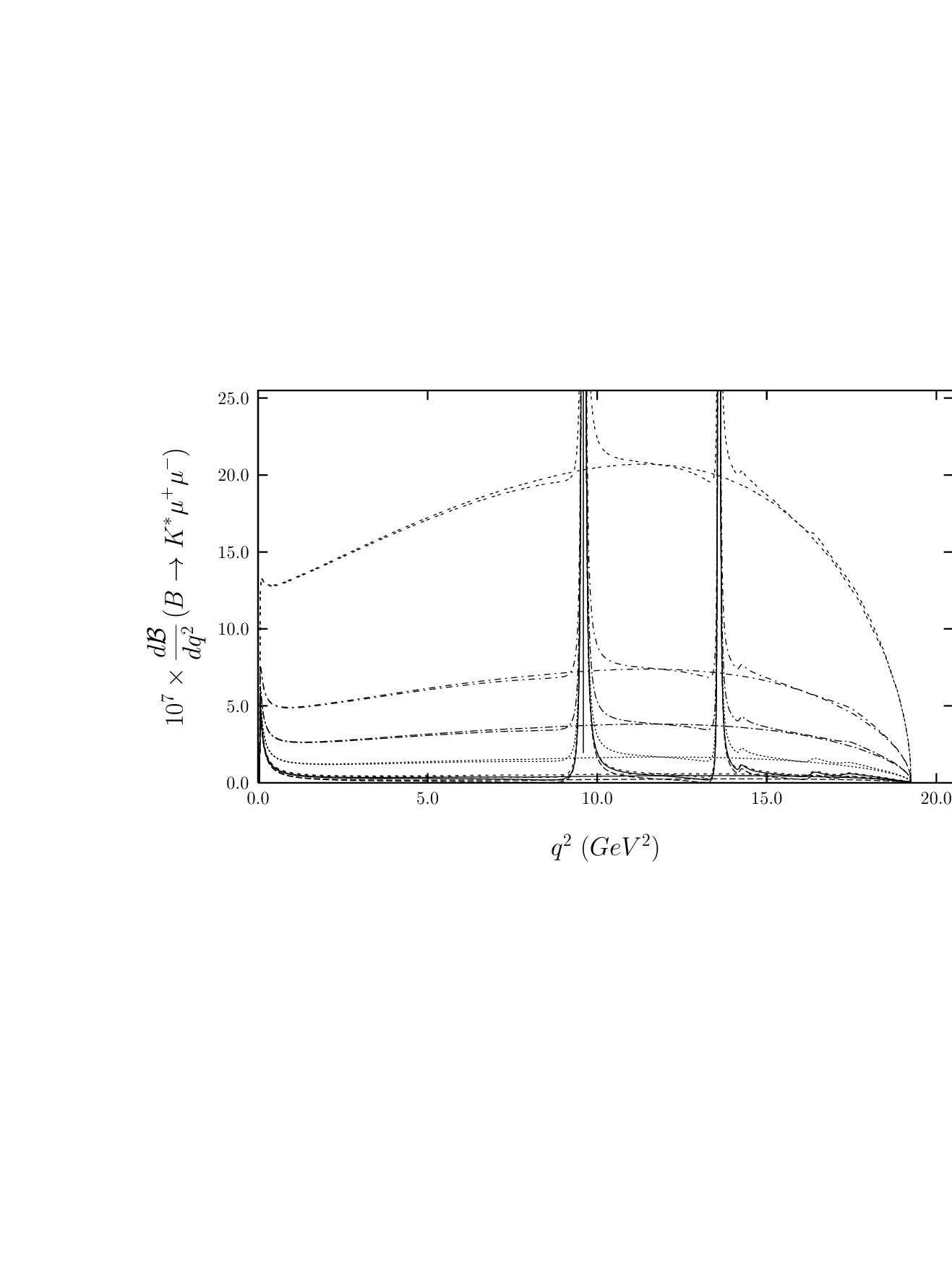}
\vskip 7.1cm
\caption{}
\end{figure}

\begin{figure}
\vskip 1.5 cm
    \includegraphics{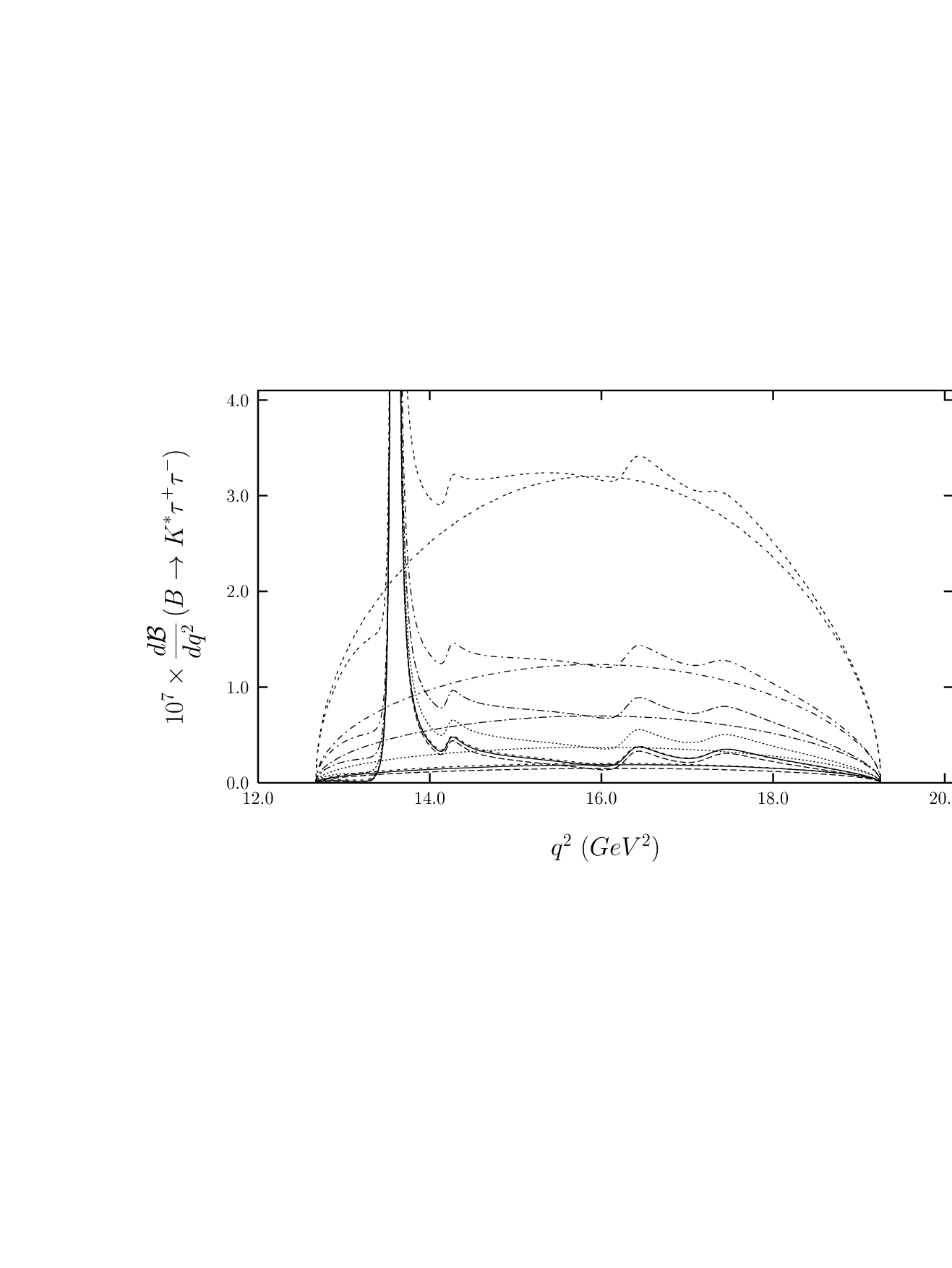}
\vskip 7.9 cm
\caption{}
\end{figure}

\begin{figure}
\vskip 1.5 cm
    \includegraphics{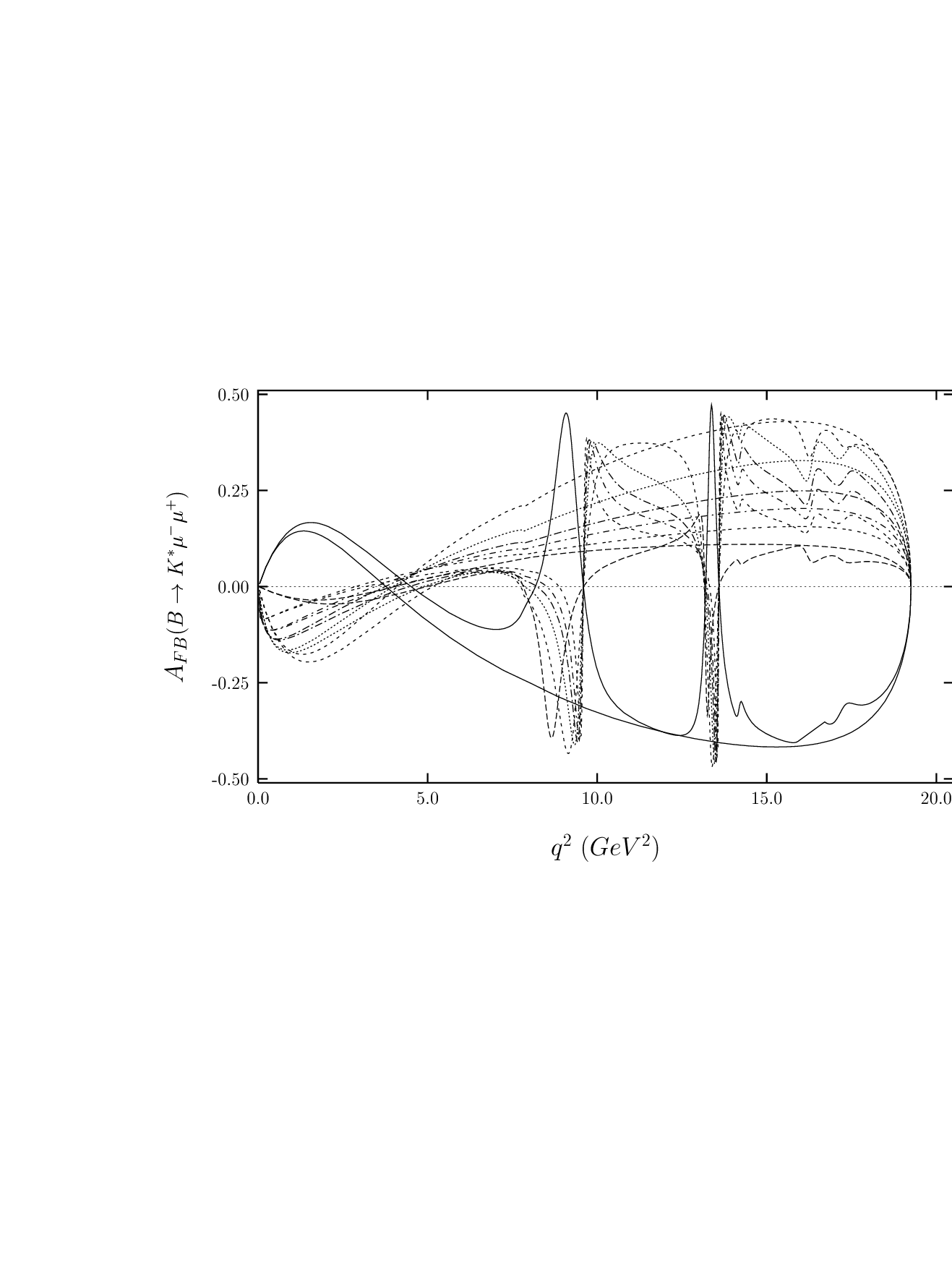}
\vskip 7.1cm
\caption{}
\end{figure}

\begin{figure}
\vskip 1.5 cm
    \includegraphics{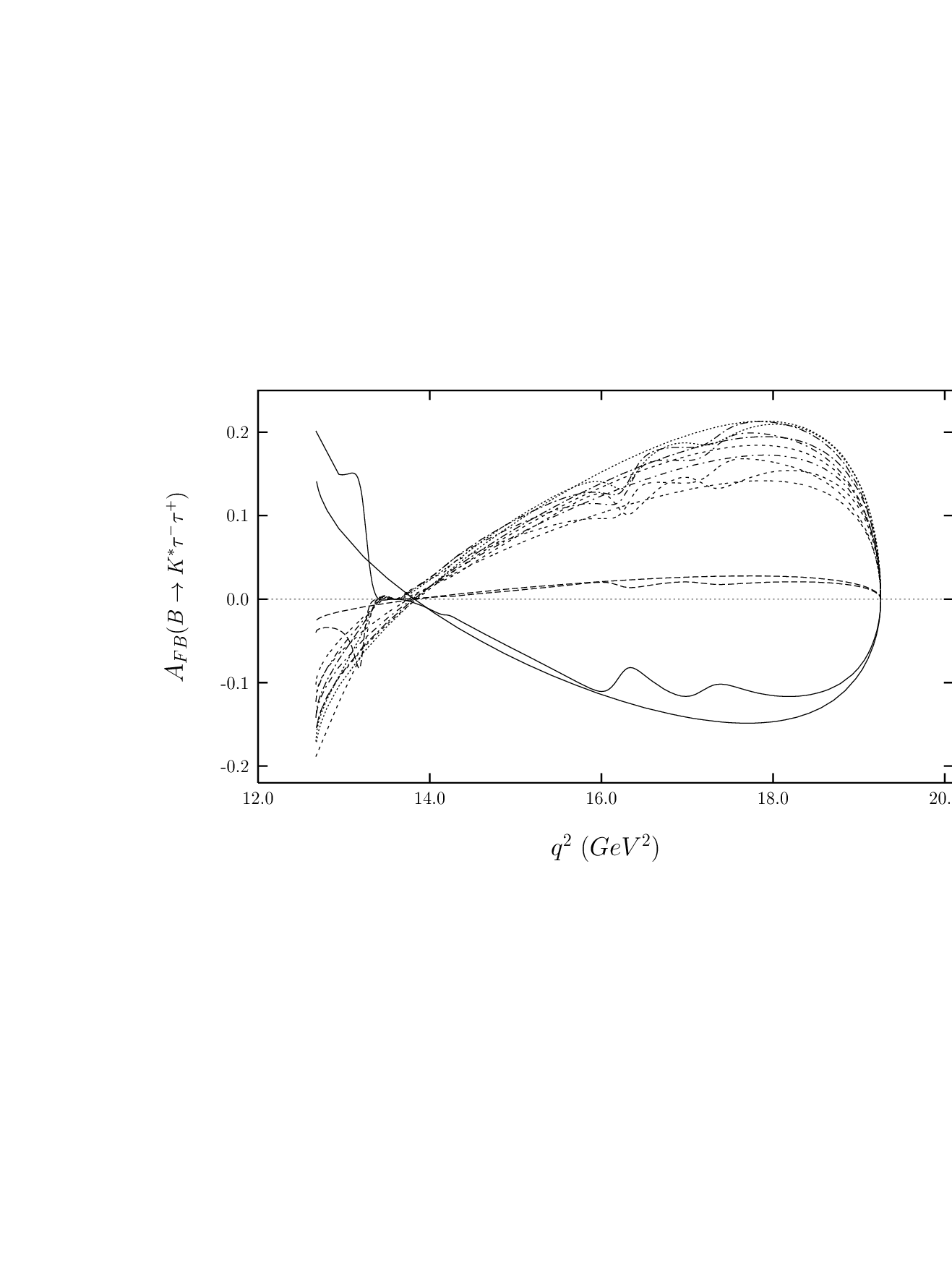}
\vskip 7.9 cm
\caption{}
\end{figure}

\begin{figure}
\vskip 1.5 cm
    \includegraphics{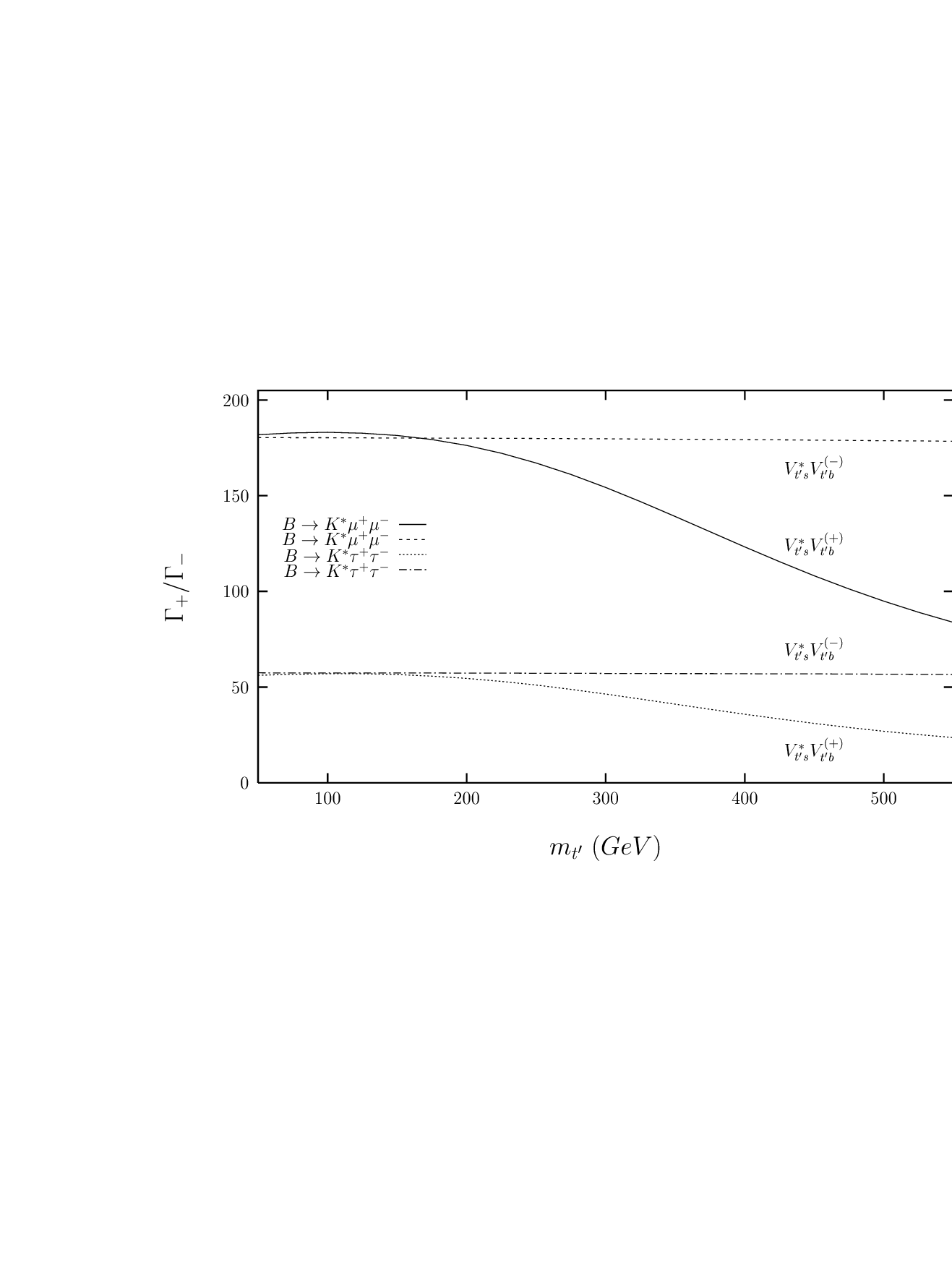}
\vskip 7.1cm
\caption{}
\end{figure}

\begin{figure}
\vskip 1.5 cm
    \includegraphics{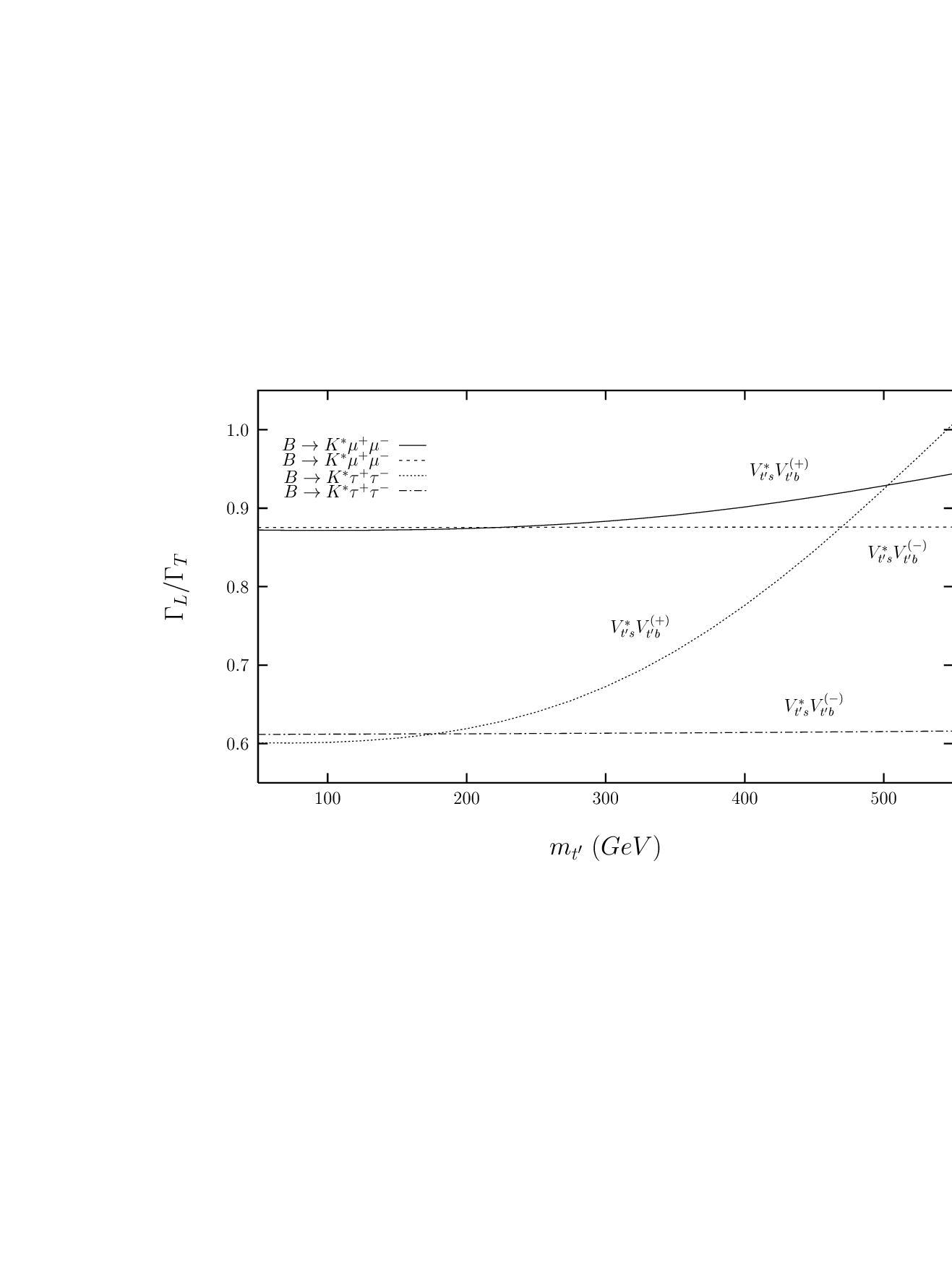}
\vskip 7.9 cm
\caption{}
\end{figure}

\end{document}